\documentclass{Interspeech}
\usepackage{times}
\usepackage{latexsym}
\usepackage{booktabs}
\usepackage{verbatim}
\usepackage[table,xcdraw]{xcolor}
\usepackage{multirow}
\usepackage[table,xcdraw]{xcolor}
\usepackage{colortbl}  
\usepackage[table,xcdraw]{xcolor}
\usepackage{booktabs}
\usepackage[T1]{fontenc}
\usepackage[utf8]{inputenc}

\usepackage{etoolbox}
\makeatletter
\patchcmd{\affiliation}
  {\renewcommand{\affiliationsep}{\vskip1pt}}
  {\renewcommand{\affiliationsep}{, }}
  {}{}
\makeatother

\usepackage{microtype}
\usepackage{inconsolata}
\usepackage{adjustbox}
\usepackage{ subcaption, algorithm2e, url, float, etoolbox, adjustbox, pgf,booktabs, verbatim}
\usepackage{xcolor}
\usepackage{amsmath}
\usepackage{amssymb}
\usepackage{tcolorbox}

\definecolor{matteSkyblue10}{RGB}{218,229,241}
\definecolor{matteSkyblue15}{RGB}{200,217,236}
\definecolor{matteSkyblue20}{RGB}{182,205,231}
\definecolor{matteSkyblue25}{RGB}{164,193,226}
\definecolor{matteSkyblue30}{RGB}{146,181,221}
\definecolor{matteSkyblue5}{RGB}{236, 239, 246} 

\title{Towards Fusion of Neural Audio Codec-based Representations with Spectral for Heart Murmur Classification via Bandit-based Cross-Attention Mechanism}

\author[affiliation={1}]{Orchid Chetia}{Phukan*}
\author[affiliation={1,2}]{Girish*}{}
\author[affiliation={1,3}]{Mohd Mujtaba}{Akhtar*}
\author[affiliation={4}]{Swarup Ranjan}{Behera}
\author[affiliation={4}]{Priyabrata}{Mallick}
\author[affiliation={5}]{Santanu}{Roy}
\author[affiliation={1}]{Arun Balaji}{Buduru}
\author[affiliation={6,7}]{Rajesh}{Sharma}

\affiliation{}{IIIT-Delhi}{India}
\affiliation{}{UPES}{India}
\affiliation{}{V.B.S.P.U}{India}
\affiliation{}{Independent Researcher}{India}
\affiliation{}{Dhineu Solutions}{India}
\affiliation{}{University of Tartu}{Estonia}
\affiliation{}{Plaksha University}{India}
\email{\textcolor{blue}{\texttt{Correspondence:}} orchidp@iiitd.ac.in} 

\usepackage{comment}

\begin{document}

\maketitle
\begingroup
  \renewcommand{\thefootnote}{\fnsymbol{footnote}}
  \setcounter{footnote}{0}
  \footnotetext{* Contributed equally as a first authors.}
\endgroup

\begin{abstract}
In this study, we focus on heart murmur classification (HMC) and hypothesize that combining neural audio codec representations (NACRs) such as EnCodec with spectral features (SFs), such as MFCC, will yield superior performance. We believe such fusion will trigger their complementary behavior as NACRs excel at capturing fine-grained acoustic patterns such as rhythm changes, spectral features focus on frequency-domain properties such as harmonic structure, spectral energy distribution crucial for analyzing the complex of heart sounds. To this end, we propose, \textbf{\texttt{BAOMI}}, a novel framework banking on novel bandit-based cross-attention mechanism for effective fusion. Here, a agent provides more weightage to most important heads in multi-head cross-attention mechanism and helps in mitigating the noise. With \textbf{\texttt{BAOMI}}, we report the topmost performance in comparison to individual NACRs, SFs, and baseline fusion techniques and setting new state-of-the-art.
\end{abstract}

\section{Introduction}
Cardiovascular diseases (CVDs) remain the leading cause of global mortality, claiming millions of lives each year~\cite{WHO2023}. Early diagnosis is vital to improving patient outcomes; however, traditional diagnostic methods, such as manual auscultation, are dependent on clinician expertise and prone to inconsistencies. Phonocardiograms (PCGs), which capture heart sounds as audio signals, offer a non-invasive, cost-effective approach for early diagnosis, identifying abnormal acoustic patterns like pathological murmurs~\cite{Dwivedi2019AlgorithmsFA}. Advances in ML have enabled the automated analysis of PCGs, enhancing diagnostic accuracy and broadening accessibility \cite{farzam2014diagnosis}. Despite these gains, challenges related to noise interference, recording variability, and the detection of subtle anomalies persist, underscoring the need for more robust methods.

In this work, we focus on Heart Murmur Classification (HMC). HMC focuses specifically on identifying and categorizing heart murmurs, which are abnormal sounds resulting from turbulent blood flow in the heart or nearby vessels. It is central to cardiac diagnostics, aiming to identify abnormal patterns within PCGs indicative of cardiovascular conditions. Early approaches relied on handcrafted spectral features, such as MFCCs and time-frequency representations, to differentiate normal and pathological heart sounds~\cite{HM1}. Such features were intially modeled using classical ML techniques such as SVM, kNN \cite{ahmad2019efficient} and later on, followed by the use of DL models such as CNN \cite{10734726} and Transformer \cite{10081829}. Further, Alkhodari et al. \cite{10081603} proposed the use of ensemble of transformer networks for HMC. Tsai  et al. \cite{tsai2023heart} proposed the use of capsule-based network that utilizes dynamic routing with MFCC features. Zhang et al. \cite{zhang2024intelligent} gave a novel parallel branch convolution and self-attention based architecture with monte-carlo dropout for HMC. \par

Recently researchers have shown the effectiveness of using neural audio codec representations (NACRs) for heart sound analysis \cite{mishra2024time}. Here, they explored the use of NACRs extract from neural audio codecs (NACs) such as EnCodec, DAC, and so on for classfying heart sounds as normal or abnormalities. Also, NACRs have shown promises in various audio and speech processing tasks such as speech separation \cite{yip24_interspeech}, environmental sound classification and various speech and audio processing applications \cite{mousavi2024dasb}. These NACRs have shown remarkable performance in these tasks as they preserve critical acoustic details while filtering out noise. However, previous works on HMC haven't focused on combining NACRs with spectral features (SFs) such as MFCC and LFCC that were traditionally used. We believe such fusion will lead to further improvement in HMC performance. So, in this study we explore such fusion for HMC and \textit{hypothesize that it will bring out their complementary strengths: NACRs excel in capturing intricate acoustic patterns such as subtle rhythm changes, while SFs emphasize frequency-domain characteristics, including harmonic structures and spectral energy distributions, which are vital for understanding the complexity of heart sounds.} To the best of our knowledge, we are the first study exploring fusion of NACRs and SFs for HMC. We introduce \texttt{\textbf{BAOMI}} (Fusion using \texttt{\textbf{BA}}ndit-based Cr\texttt{\textbf{O}}ss-Attention \texttt{\textbf{M}}echan\texttt{\textbf{I}}sm), a novel framework for effective fusion of NACRs and SFs, which utilizes a novel bandit-based cross-attention mechanism. \texttt{\textbf{BAOMI}} employs an agent to identify the most relevant attention heads and providing them more weightage within the multi-head cross-attention mechanism, reducing the impact of noise. Using \texttt{\textbf{BAOMI}}, we achieve superior performance compared to individual NACRs and SFs, strong baseline fusion methods and we establish state-of-the-art (SOTA) results against existing methods from previous works. \par

\noindent \textbf{To summarize, the main contributions of the paper are:} \par
\begin{itemize}
    \item We propose, \texttt{\textbf{BAOMI}} for effective fusion of NACRs with SFs. It introduces a novel bandit-based cross-attention mechanism that dynamically identifies and utilizes the most effective attention heads for fusion.

    \item With \texttt{\textbf{BAOMI}}, we achieve superior results compared to models using NACRs or SFs individually, baseline fusion techniques and establishes new SOTA performance for HMC, outperforming previous methods. 

    \item The superior performance showed by the combination of NACRs and SFs through \texttt{\textbf{BAOMI}} can be attributed to the emergence of complementary behavior by NACRs and SFs where NACRs focuses on the fine-grained acoustic details while SFs focuses on frequency-domain properties critical for heart sound analysis.
\end{itemize}

\noindent The source code and trained model checkpoints for this work can be found at: \url{https://github.com/Helix-IIIT-Delhi/BAOMI-Heart_Murmur}

\begin{figure}[!bt]
    \centering
    \includegraphics[width=0.8\linewidth]{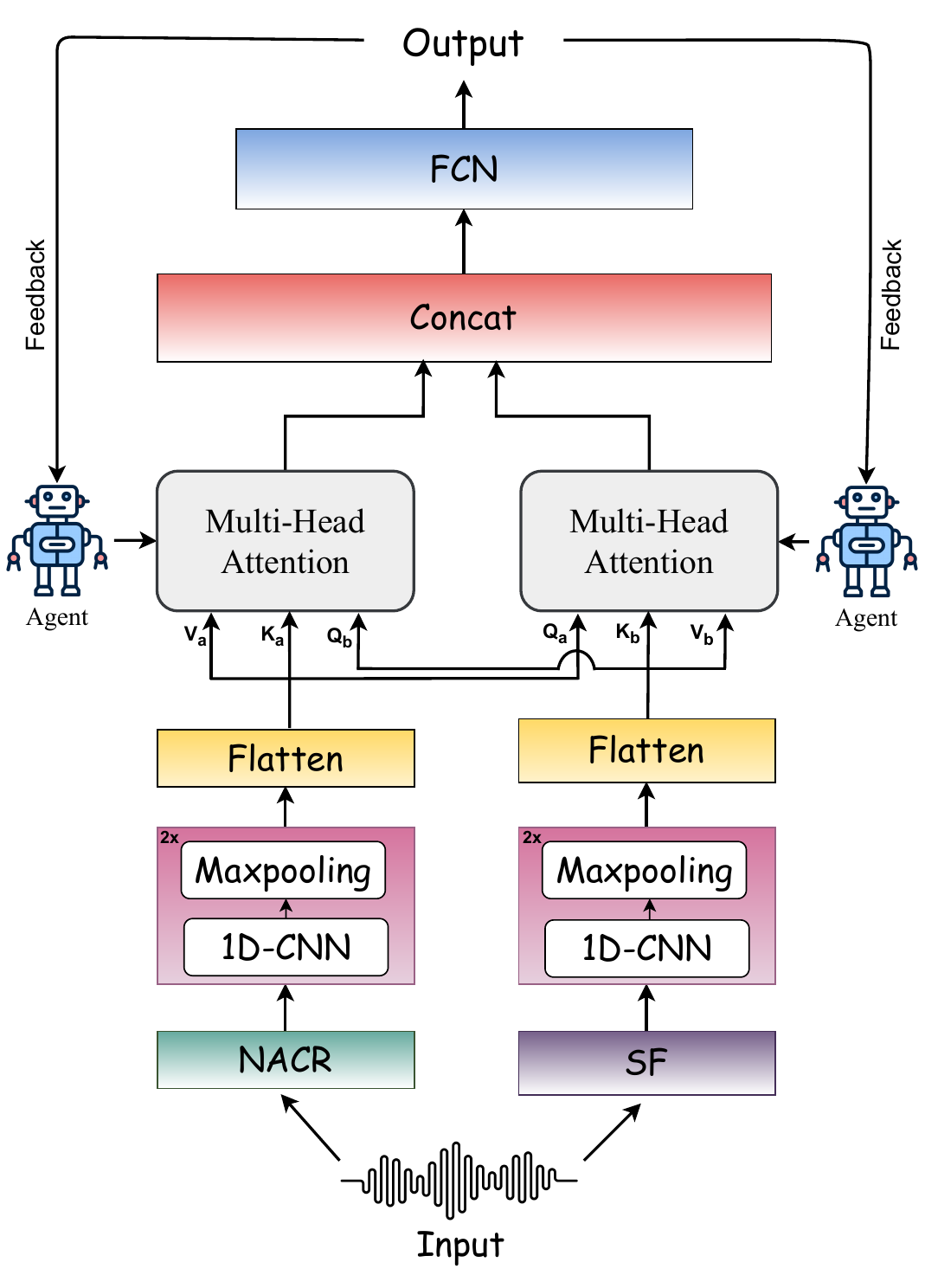}
    \caption{Proposed Framework: \textbf{\texttt{BAOMI}}}
    \label{fig:proposed}
\end{figure}

\section{Neural Audio Codec}

In this section, we discuss the SOTA NACs considered in our study. 

\noindent \textbf{EnCodec} \cite{fossez2023high}: It generates compact, quantized audio representations through a streaming encoder-decoder architecture utilizing RVQ. By incorporating multi-scale spectrogram adversarial losses and Transformer-based entropy coding, EnCodec strikes a balance between compression efficiency and audio quality.
Its bandwidth ranges from 1.5 to 24 kbps with sample rates up to 48 kHz. We use the EnCodec versions that requires that input to resample to 24 kHz\footnote{\url{https://huggingface.co/facebook/encodec_24khz}} (EnCodec24) and 48 (EnCodec48) kHz\footnote{\url{https://huggingface.co/facebook/encodec_48khz}}. \par
\noindent \textbf{Speech Tokenizer}\footnote{\url{https://github.com/ZhangXInFD/SpeechTokenizer}} \cite{zhang2024speechtokenizer}: It is optimized for speech language models, merging semantic and acoustic tokens into a unified representation. Utilizing a RVQ-based encoder-decoder framework, its hierarchical design encodes semantic content in early layers, followed by acoustic features in later layers. It offers reconstruction quality comparable to EnCodec. \par
\noindent \textbf{Descript Audio Codec (DAC)}\footnote{\url{https://github.com/descriptinc/descript-audio-codec}} \cite{kumar2024high}: It provides highly efficient audio compression, achieving remarkable quality retention. It compresses 44.1 kHz audio up to 90 times into discrete tokens at 8 kbps. It employs multi-period and multi-band Short-Time Fourier Transform (STFT) discriminators to minimize high-frequency artifacts, while multi-scale STFT and mel-spectral loss functions enhance fidelity.  

\noindent \textbf{SNAC}\footnote{\url{https://github.com/hubertsiuzdak/snac}} \cite{siuzdak2024snac}: It introduces a multi-scale approach to audio compression, employing RVQ across different temporal resolutions. This layered quantization captures both fine and broad audio details, enhancing perceptual quality with local windowed attention. We use the SNAC versions that requires input audios to be sampled 24 kHz (SNAC24) and 32 kHz (SNAC32). \par
\noindent Before processing by DAC and Speech Tokenizer, all input audio signals undergo resampling to 16 kHz. The frozen encoders of Encodec, DAC, are then used to extract NACRs, which are obtained through mean pooling. For Speech Tokenizer, we add the codes and average it to get a single vector representation and for SNAC, we concat the codes extracted as these codes were already in single-dimension vector, so no further average pooling is required. Finally, we receive NACRs in feature dimensions of 3225 for DAC, and 3226 for Speech Tokenizer. Encodec 24 kHz, Encodec 48 KhZ gives 4839 and 150-size NACRs, whereas, SNAC 24 kHz and SNAC 32 kHZ gives 5292 and 10080-size NACRs.\par

\section{Modeling}

In this section, we explain the SFs considered, the downstream modeling for individual features, and the proposed framework, \texttt{\textbf{BAOMI}} for fusion of NACRs with SFs. \par

\noindent\textbf{Spectral Features}: As SFs, we consider mel-frequency cepstral coefficients (MFCC)\footnote{\url{https://librosa.org/doc/main/generated/librosa.feature.mfcc.html}} and linear-frequency cepstral coefﬁcients (LFCC)\footnote{\url{https://spafe.readthedocs.io/en/latest/features/lfcc.html}}. We use the default parameters as given in the library and get 14-dimension LFCC in return. For MFCC, we set number of MFCCs to return as 40. \par

\noindent\textbf{Downstream Modeling}: We use Fully Connected Network (FCN) and CNN as downstream network with individual NACRs and SFs. The CNN framework consists of two 1D convolutional layers with 64 and 128 filters, respectively, and a kernel size of 3 with ReLu as the activation function. We use maxpooling after each 1D convolutional layer. The output is flattened and routed through a dense layer with 128 neurons and ReLU activation, followed by a softmax activation function for classification. In a similar manner, the FCN flattens the input features by first activating a dense layer of 256 neurons through ReLU. We then connect the output layer for classification with softmax activation function. 

\subsection{BAOMI: Bandit-based Cross-Attention Mechanism}

We propose, \textbf{\texttt{BAOMI}} for the fusion of NACRs with SFs. The proposed framework is illustrated in Figure \ref{fig:proposed}. \textbf{\texttt{BAOMI}} applies a bandit-based multi-head cross-attention mechanism that dynamically adjusts attention head importance based on their contribution to task-specific loss reduction. By leveraging a multi-armed bandit strategy, \textbf{\texttt{BAOMI}} learns to prioritize the most informative heads, ensuring optimal fusion and mitigating the noise induced by heads. First the representations are passed through two 1D convolutional layers and we keep the modeling details as same as used in the downstream modeling with individual features above. Following this, the features are flattened and passed through the bandit-based cross-attention mechanism. The set of steps following this are detailed below in details:

\noindent\textbf{Multi-Head Cross-Attention}: Let (\(\mathbf{Z}_A\)) and (\(\mathbf{Z}_B\)) be the flattened features corresponding to NACR ($\mathbf{A}$) and SF ($\mathbf{B}$) networks. We first project them into multi-head query ($\mathbf{Q}$), key ($\mathbf{K}$), and value ($\mathbf{V}$) representations for \( H \) heads:

\begin{align}
    \mathbf{Q}_h^A &= \mathbf{Z}_A \mathbf{W}_{Q,h}, \quad
    \mathbf{K}_h^B = \mathbf{Z}_B \mathbf{W}_{K,h}, \quad
    \mathbf{V}_h^B = \mathbf{Z}_B \mathbf{W}_{V,h} \\
    \mathbf{Q}_h^B &= \mathbf{Z}_B \mathbf{W}_{Q,h}, \quad
    \mathbf{K}_h^A = \mathbf{Z}_A \mathbf{W}_{K,h}, \quad
    \mathbf{V}_h^A = \mathbf{Z}_A \mathbf{W}_{V,h}
\end{align}

\noindent where \(\mathbf{W}_{Q,h}, \mathbf{W}_{K,h}, \mathbf{W}_{V,h}\) are learnable weight matrices for head \( h \). The cross-attention computation per head is:

\begin{align}
    \mathbf{A}_h^{A \rightarrow B} &= \text{softmax} \left( \frac{\mathbf{Q}_h^A (\mathbf{K}_h^B)^\top}{\sqrt{d_h}} \right) \mathbf{V}_h^B \\
    \mathbf{A}_h^{B \rightarrow A} &= \text{softmax} \left( \frac{\mathbf{Q}_h^B (\mathbf{K}_h^A)^\top}{\sqrt{d_h}} \right) \mathbf{V}_h^A
\end{align}

\noindent where \( d_h \) is the dimension of head. Each head captures cross-feature interactions independently.

\noindent\textbf{Bandit-based Head Weightage}: We employ a multi-armed bandit approach to dynamically learn head importance at the head-level. This step is being calclulated for both NACR ($\mathbf{A}$) and SF ($\mathbf{B}$) networks. Each head maintains a Q-value \( Q_h \) that tracks its performance in reducing the task-specific loss i.e. cross-entropy for HMC. The Q-value update is defined as: $ Q_h (t+1) = \gamma Q_h (t) + (1 - \gamma) R_h $  where, \( Q_h (t) \) is the past Q-value of head \( h \), \( \gamma \) is the reward decay factor, \( R_h \) is the reward for head \( h \) and computed as: 

\begin{equation}
R_h = \frac{\Delta L_h}{\sum_{h'} \Delta L_{h'} + \epsilon} 
\end{equation} 
\noindent where \( \Delta L_h = L_{\text{prev}} - L_{\text{curr}} \) quantifies the reduction in task-specific loss. Heads that contribute more to loss reduction receive higher rewards.

\noindent \textbf{Soft-Head Weighting for Fusion}: We compute a soft-weighted combination of all heads using the learned Q-values for both NACR ($\mathbf{A}$) and SF ($\mathbf{B}$) networks:

\begin{equation}
    W_h = \frac{\exp(Q_h)}{\sum_{h'} \exp(Q_{h'})}
\end{equation}

\noindent The final fused representation is obtained by concatenating the weighted cross-attention outputs of NACR ($\mathbf{A}$) and SF ($\mathbf{B}$) networks:

\begin{equation}
\begin{aligned}
\mathbf{Z}_{\text{fused}} = \text{Concat} \Bigg( 
& \sum_{h=1}^{H} W_h \cdot \mathbf{A}_h^{A \rightarrow B}, 
& \sum_{h=1}^{H} W_h \cdot \mathbf{A}_h^{B \rightarrow A} 
\Bigg)
\end{aligned}
\end{equation}

\noindent This ensures that the most informative heads from both attention flows contribute effectively to the final representation. We append a fully connected network (FCN) block on top of $\mathbf{Z}_{\text{fused}}$ with a dense layer of 128 neurons and an output layer with a softmax activation function for HMC. We set the number of heads to be 4. The trainable parameters range from 11.8M to 12.6M, depending on the size of the input representation.


\section{Experiments}

\subsection{Dataset}

\noindent We use CirCor DigiScope dataset \cite{oliveira2021circor}, available on PhysioNet \cite{goldberger2000physiobank}. We specifically worked with the publicly accessible subset containing data from 963 patients. The dataset consists of three class labels: Present, Absent, and Unknown, with a distribution of 179, 695, and 68 samples, respectively, indicating an inherent class imbalance. Each sample comprises multiple variable-length PCG recordings, totaling 3,163 recordings with recording durations ranging from 5 to 65 seconds. We only work with the Present and Absent class due to very less number of samples present in the Unknown class. As the NACs give variable NACRs for different lengths of input audio, so to prevent this, we pad the audios to the length of the maximum duration audio. 

\noindent\textbf{Training and Hyperparameter Details}: We keep the batch size as 32 and train the models for 50 epochs. We use Adam as optimizer with cross-entropy as the loss function. We use five-fold cross validation for training and testing where four folds are used for training and one fold for testing. 

\subsection{Experimental Results}

We present the results of our experiments with individual NACRs and SFs for HMC with downstream models in Table \ref{tab:single}. We use Accuracy, Macro Average F1-score (MA-F1), and Weighted Average F1-score (WA-F1) as the evaluation metrics. From the results, it is evident that LFCC and MFCC consistently outperform NACRs across both FCN and CNN downstream with CNN models showing relatively better performance. MFCC features achieves the highest performance across all metrics. Similarly, LFCC attains the second-highest scores, reinforcing the effectiveness of SFs for HMC \cite{tsai2023heart, das2024heart}. In contrast, NACRs exhibit lower performance. Among the NACRs, SNAC24 achieves the highest MA-F1, while DAC performs best in terms of accuracy and WA-F1. These mixed results points that the performance changes depending on the downstream data distribution and also how each NAC was pre-trained and the data during its pre-training. So, we can say that SNAC24 and DAC has better implicit transferability for HMC. Despite being the best among NACRs, both SNAC24 and DAC fall short of the performance achieved by SFs, LFCC and MFCC. The superior performance of SFs can be attributed to their ability to emphasize frequency-domain characteristics, including harmonic structures and spectral energy distributions, which are vital for understanding the complexity of heart sounds. These spectral features retain rich discriminative information, making them more effective than NACRs for HMC. \par

In Table \ref{tab:fusion}, we present the results obtained with different combinations. We use cross-attention as a baseline for validation our proposed novel method, \textbf{\texttt{BAOMI}}. Cross-Attention is one of the most preferred fusion technique by previous researchers for feature fusion \cite{ilias24_interspeech, despotovic24_interspeech} and this makes cross-attention a strong baseline. For fair comparison, we keep the modeling same as used for \textbf{\texttt{BAOMI}} by discarding the novel Bandit-based head weightage mechanism. We keep the training details same as \textbf{\texttt{BAOMI}}. The results demonstrate the effectiveness of \textbf{\texttt{BAOMI}}, which consistently outperforms the Cross-Attention baseline across all combinations. This validates the contribution of the Bandit-based head weightage mechanism in improving fusion by removing noise injected through unimportant attention heads in the multi-head cross attention mechanism. By dynamically adjusting the contribution of different attention heads based on the input, \textbf{\texttt{BAOMI}} ensures that only the most relevant information is retained, leading to more effective feature integration. With \textbf{\texttt{BAOMI}}, through the fusion of DAC and MFCC, we got the best performance amongst all the combinations and thus showing its effective complementary strength as DAC showed top within NACRs and MFCC within SFs individually for HMC. Overall, we observe that the fusion of NACRs and SFs generally lead to improved performance in comparison to homogenous fusion of NACRs and SFs. This validates \textit{our hypothesis that fusion of NACRs and SFs will be the most effective for HMC due to emergence of complementary behavior amongst them}. Further, we can also see that fusion of SFs leads to improve performance than their individual modeling. We have also plotted the t-SNE plots visualizations from representations extracted from the penultimate layers of the models with MFCC, LFCC, \textbf{\texttt{BAOMI}} with the fusion of DAC + MFCC and SNAC24 + LFCC (another best pair). We observe better clustering across the classes with the \textbf{\texttt{BAOMI}} models. This supports our results obtained. We also plot the confusion matrix of \textbf{\texttt{BAOMI}} with the fusion of DAC + MFCC and SNAC24 + LFCC. As for validating the effectiveness of our proposed methods in comparison to previous works, we already showed in our experiments in Table \ref{tab:fusion}, that fusion of DAC (NACR) and MFCC (SF) through \textbf{\texttt{BAOMI}} improves over individual SFs (See in Table \ref{tab:single}) which has shown SOTA performance in previous works \cite{deng2020heart, kui2021heart, li2022heart, tsai2023heart} for heart sound classification. This shows that the proposed approach sets new SOTA for HMC.

\begin{table}[!bt]
\setlength{\tabcolsep}{6pt}
\scriptsize
\centering
\begin{tabular}{lcccccc}
\toprule
\multicolumn{1}{c|}{\multirow{2}{*}{\textbf{R}}} & \multicolumn{3}{c|}{\textbf{FCN}}                                     & \multicolumn{3}{c}{\textbf{CNN}}                 \\ 
\cmidrule(lr){2-4} \cmidrule(lr){5-7}
\multicolumn{1}{c|}{}                            & \textbf{Acc}   & \textbf{MA-F1} & \multicolumn{1}{c|}{\textbf{WA-F1}} & \textbf{Acc}   & \textbf{MA-F1} & \textbf{WA-F1} \\ 
\midrule
\multicolumn{7}{c}{\textbf{NACRs}} \\ 
\midrule
\multicolumn{1}{l|}{E24} & \cellcolor{matteSkyblue15}70.30 & \cellcolor{matteSkyblue5}49.25 & \multicolumn{1}{c|}{\cellcolor{matteSkyblue10}66.11} & \cellcolor{matteSkyblue15}74.42 & \cellcolor{matteSkyblue10}50.60 & \cellcolor{matteSkyblue15}68.61 \\
\multicolumn{1}{l|}{E48} & \cellcolor{matteSkyblue10}68.23 & \cellcolor{matteSkyblue5}45.03 & \multicolumn{1}{c|}{\cellcolor{matteSkyblue10}64.78} & \cellcolor{matteSkyblue15}70.60 & \cellcolor{matteSkyblue5}47.97 & \cellcolor{matteSkyblue10}65.98 \\
\multicolumn{1}{l|}{D}   & \cellcolor{matteSkyblue15}72.07 & \cellcolor{matteSkyblue15}52.31 & \multicolumn{1}{c|}{\cellcolor{matteSkyblue15}69.42} & \cellcolor{matteSkyblue20}75.75 & \cellcolor{matteSkyblue15}53.17 & \cellcolor{matteSkyblue15}70.24 \\
\multicolumn{1}{l|}{ST}  & \cellcolor{matteSkyblue15}71.12 & \cellcolor{matteSkyblue5}43.14 & \multicolumn{1}{c|}{\cellcolor{matteSkyblue10}65.78} & \cellcolor{matteSkyblue20}75.75 & \cellcolor{matteSkyblue5}44.41 & \cellcolor{matteSkyblue15}66.32 \\
\multicolumn{1}{l|}{S24} & \cellcolor{matteSkyblue10}68.56 & \cellcolor{matteSkyblue15}64.96 & \multicolumn{1}{c|}{\cellcolor{matteSkyblue10}65.36} & \cellcolor{matteSkyblue25}76.41 & \cellcolor{matteSkyblue20}69.88 & \multicolumn{1}{c}{\cellcolor{matteSkyblue20}69.02} \\
\multicolumn{1}{l|}{S32} & \cellcolor{matteSkyblue5}59.46  & \cellcolor{matteSkyblue15}63.59 & \multicolumn{1}{c|}{\cellcolor{matteSkyblue5}60.74}  & \cellcolor{matteSkyblue15}71.10 & \cellcolor{matteSkyblue15}68.65 & \cellcolor{matteSkyblue15}66.47 \\ 
\midrule
\multicolumn{7}{c}{\textbf{SFs}} \\ 
\midrule
\multicolumn{1}{l|}{\textbf{L}} & \cellcolor{matteSkyblue20}\textbf{75.14} & \cellcolor{matteSkyblue20}\textbf{69.85} & \multicolumn{1}{c|}{\cellcolor{matteSkyblue20}\textbf{71.54}} & \cellcolor{matteSkyblue25}\textbf{79.73} & \cellcolor{matteSkyblue25}\textbf{69.87} & \multicolumn{1}{c}{\cellcolor{matteSkyblue25}\textbf{75.13}} \\
\multicolumn{1}{l|}{\textbf{M}} & \cellcolor{matteSkyblue25}\textbf{78.56} & \cellcolor{matteSkyblue20}\textbf{70.68} & \multicolumn{1}{c|}{\cellcolor{matteSkyblue25}\textbf{76.63}} & \cellcolor{matteSkyblue25}\textbf{80.90} & \cellcolor{matteSkyblue25}\textbf{73.28} & \multicolumn{1}{c}{\cellcolor{matteSkyblue25}\textbf{77.63}} \\ 
\bottomrule
\end{tabular}
\caption{Performance Scores of different NACRs and SFs with FCN and CNN downstream; Abbreviations used: Macro Average F1 (MA-F1), Weighted Average F1 (WA-F1), E24 (EnCodec24), E48 (EnCodec48), D (DAC), ST (Speech Tokenizer), S24 (SNAC24), S32 (SNAC32), L (LFCC), and M (MFCC); The scores are presented in \% and are the average of five folds; Abbreviations used in this Table are kept same for Table \ref{tab:fusion}}
\label{tab:single}
\end{table}

\begin{table}[!ht]
\setlength{\tabcolsep}{5pt}
\scriptsize
\centering
\begin{tabular}{lccc|ccc}
\toprule
\multicolumn{1}{l|}{\textbf{Pairs}} & \multicolumn{3}{c|}{\textbf{Cross-Attention}} & \multicolumn{3}{c}{\textbf{BAOMI}} \\
\cmidrule(lr){2-4} \cmidrule(lr){5-7}
\multicolumn{1}{l|}{} 
    & \textbf{Acc} & \textbf{MA-F1} & \textbf{WA-F1} 
    & \textbf{Acc} & \textbf{MA-F1} & \textbf{WA-F1} \\ 
\midrule
\multicolumn{7}{c}{\textbf{NACRs + NACRs}} \\ 
\midrule
\multicolumn{1}{l|}{E24 + D} 
    & \cellcolor{matteSkyblue20}77.43 
    & \cellcolor{matteSkyblue10}50.59 
    & \cellcolor{matteSkyblue25}72.55 
    & \cellcolor{matteSkyblue25}81.14 
    & \cellcolor{matteSkyblue15}54.08 
    & \cellcolor{matteSkyblue25}73.11 \\
\multicolumn{1}{l|}{E24 + ST} 
    & \cellcolor{matteSkyblue25}80.48 
    & \cellcolor{matteSkyblue15}59.88 
    & \cellcolor{matteSkyblue25}77.31 
    & \cellcolor{matteSkyblue25}83.20 
    & \cellcolor{matteSkyblue15}62.43 
    & \cellcolor{matteSkyblue25}79.03 \\
\multicolumn{1}{l|}{E24 + S24} 
    & \cellcolor{matteSkyblue25}79.12 
    & \cellcolor{matteSkyblue10}53.64 
    & \cellcolor{matteSkyblue25}74.40 
    & \cellcolor{matteSkyblue25}79.81 
    & \cellcolor{matteSkyblue10}54.17 
    & \cellcolor{matteSkyblue25}75.02 \\
\multicolumn{1}{l|}{E24 + S32} 
    & \cellcolor{matteSkyblue25}81.16 
    & \cellcolor{matteSkyblue10}51.20 
    & \cellcolor{matteSkyblue25}72.54 
    & \cellcolor{matteSkyblue25}81.92 
    & \cellcolor{matteSkyblue10}51.74 
    & \cellcolor{matteSkyblue25}73.20 \\
\multicolumn{1}{l|}{E48 + D} 
    & \cellcolor{matteSkyblue25}79.47 
    & \cellcolor{matteSkyblue10}53.37 
    & \cellcolor{matteSkyblue25}74.44 
    & \cellcolor{matteSkyblue25}81.24 
    & \cellcolor{matteSkyblue10}53.90 
    & \cellcolor{matteSkyblue25}76.11 \\
\multicolumn{1}{l|}{E48 + ST} 
    & \cellcolor{matteSkyblue25}82.01 
    & \cellcolor{matteSkyblue10}52.11 
    & \cellcolor{matteSkyblue25}74.95 
    & \cellcolor{matteSkyblue25}82.81 
    & \cellcolor{matteSkyblue10}52.63 
    & \cellcolor{matteSkyblue25}75.67 \\
\multicolumn{1}{l|}{E48 + S24} 
    & \cellcolor{matteSkyblue25}80.48 
    & \cellcolor{matteSkyblue10}55.56 
    & \cellcolor{matteSkyblue25}75.67 
    & \cellcolor{matteSkyblue25}83.20 
    & \cellcolor{matteSkyblue15}58.11 
    & \cellcolor{matteSkyblue25}76.40 \\
\multicolumn{1}{l|}{E48 + S32} 
    & \cellcolor{matteSkyblue25}79.80 
    & \cellcolor{matteSkyblue10}50.80 
    & \cellcolor{matteSkyblue25}73.13 
    & \cellcolor{matteSkyblue25}79.89 
    & \cellcolor{matteSkyblue15}58.55 
    & \cellcolor{matteSkyblue25}74.31 \\
\multicolumn{1}{l|}{D + ST} 
    & \cellcolor{matteSkyblue20}78.96 
    & \cellcolor{matteSkyblue15}59.94 
    & \cellcolor{matteSkyblue25}76.59 
    & \cellcolor{matteSkyblue25}80.82 
    & \cellcolor{matteSkyblue15}60.80 
    & \cellcolor{matteSkyblue25}78.19 \\
\multicolumn{1}{l|}{D + S24} 
    & \cellcolor{matteSkyblue20}77.88 
    & \cellcolor{matteSkyblue10}53.20 
    & \cellcolor{matteSkyblue25}73.96 
    & \cellcolor{matteSkyblue25}79.81 
    & \cellcolor{matteSkyblue10}56.06 
    & \cellcolor{matteSkyblue25}75.46 \\
\multicolumn{1}{l|}{D + S32} 
    & \cellcolor{matteSkyblue20}77.57 
    & \cellcolor{matteSkyblue10}52.18 
    & \cellcolor{matteSkyblue25}74.50 
    & \cellcolor{matteSkyblue25}81.32 
    & \cellcolor{matteSkyblue10}58.53 
    & \cellcolor{matteSkyblue25}77.14 \\
\multicolumn{1}{l|}{ST + S24} 
    & \cellcolor{matteSkyblue25}81.32 
    & \cellcolor{matteSkyblue15}58.53 
    & \cellcolor{matteSkyblue25}77.14 
    & \cellcolor{matteSkyblue25}83.81 
    & \cellcolor{matteSkyblue15}59.37 
    & \cellcolor{matteSkyblue25}77.67 \\
\multicolumn{1}{l|}{ST + S32} 
    & \cellcolor{matteSkyblue25}81.83 
    & \cellcolor{matteSkyblue15}59.37 
    & \cellcolor{matteSkyblue25}77.67 
    & \cellcolor{matteSkyblue25}82.92 
    & \cellcolor{matteSkyblue15}59.37 
    & \cellcolor{matteSkyblue25}78.51 \\ 
\midrule
\multicolumn{7}{c}{\textbf{SFs + SFs}} \\ 
\midrule
\multicolumn{1}{l|}{L + M} 
    & \cellcolor{matteSkyblue25}83.55 
    & \cellcolor{matteSkyblue25}68.10 
    & \cellcolor{matteSkyblue25}81.23 
    & \cellcolor{matteSkyblue25}85.24 
    & \cellcolor{matteSkyblue20}71.51 
    & \cellcolor{matteSkyblue25}85.29 \\ 
\midrule
\multicolumn{7}{c}{\textbf{NACRs + SFs}} \\ 
\midrule
\multicolumn{1}{l|}{E24 + L} 
    & \cellcolor{matteSkyblue25}83.96 
    & \cellcolor{matteSkyblue20}71.09 
    & \cellcolor{matteSkyblue25}82.15 
    & \cellcolor{matteSkyblue25}85.73 
    & \cellcolor{matteSkyblue25}75.62 
    & \cellcolor{matteSkyblue25}85.80 \\
\multicolumn{1}{l|}{E24 + M} 
    & \cellcolor{matteSkyblue25}82.99 
    & \cellcolor{matteSkyblue20}70.19 
    & \cellcolor{matteSkyblue25}80.70 
    & \cellcolor{matteSkyblue25}86.32 
    & \cellcolor{matteSkyblue25}78.53 
    & \cellcolor{matteSkyblue25}85.14 \\
\multicolumn{1}{l|}{E48 + L} 
    & \cellcolor{matteSkyblue25}82.29 
    & \cellcolor{matteSkyblue20}70.55 
    & \cellcolor{matteSkyblue25}84.31 
    & \cellcolor{matteSkyblue25}85.44 
    & \cellcolor{matteSkyblue25}75.45 
    & \cellcolor{matteSkyblue25}84.89 \\
\multicolumn{1}{l|}{E48 + M} 
    & \cellcolor{matteSkyblue25}83.82 
    & \cellcolor{matteSkyblue20}68.80 
    & \cellcolor{matteSkyblue25}82.19 
    & \cellcolor{matteSkyblue25}85.34 
    & \cellcolor{matteSkyblue25}75.45 
    & \cellcolor{matteSkyblue25}86.89 \\
\multicolumn{1}{l|}{D + L} 
    & \cellcolor{matteSkyblue25}85.73 
    & \cellcolor{matteSkyblue20}72.18 
    & \cellcolor{matteSkyblue25}84.46 
    & \cellcolor{matteSkyblue25}87.29 
    & \cellcolor{matteSkyblue25}78.55 
    & \cellcolor{matteSkyblue25}89.31 \\
\multicolumn{1}{l|}{\textbf{D + M}}  
    & \cellcolor{matteSkyblue25}\textbf{89.81} 
    & \cellcolor{matteSkyblue20}\textbf{75.06} 
    & \cellcolor{matteSkyblue25}\textbf{85.49} 
    & \cellcolor{matteSkyblue25}\textbf{89.93} 
    & \cellcolor{matteSkyblue25}\textbf{79.37} 
    & \cellcolor{matteSkyblue25}\textbf{89.67} \\
\multicolumn{1}{l|}{ST + L} 
    & \cellcolor{matteSkyblue25}84.19 
    & \cellcolor{matteSkyblue20}74.08 
    & \cellcolor{matteSkyblue25}85.41 
    & \cellcolor{matteSkyblue25}86.82 
    & \cellcolor{matteSkyblue25}78.80 
    & \cellcolor{matteSkyblue25}87.59 \\
\multicolumn{1}{l|}{ST + M} 
    & \cellcolor{matteSkyblue25}86.47 
    & \cellcolor{matteSkyblue20}74.85 
    & \cellcolor{matteSkyblue25}85.44 
    & \cellcolor{matteSkyblue25}88.73 
    & \cellcolor{matteSkyblue25}77.38 
    & \cellcolor{matteSkyblue25}88.63 \\
\multicolumn{1}{l|}{S24 + L} 
    & \cellcolor{matteSkyblue25}88.96 
    & \cellcolor{matteSkyblue20}71.60 
    & \cellcolor{matteSkyblue25}80.21 
    & \cellcolor{matteSkyblue25}89.08 
    & \cellcolor{matteSkyblue25}75.26 
    & \cellcolor{matteSkyblue25}85.99 \\
\multicolumn{1}{l|}{S24 + M} 
    & \cellcolor{matteSkyblue25}88.32 
    & \cellcolor{matteSkyblue20}74.18 
    & \cellcolor{matteSkyblue25}81.50 
    & \cellcolor{matteSkyblue25}89.79 
    & \cellcolor{matteSkyblue25}75.91 
    & \cellcolor{matteSkyblue25}86.60 \\
\multicolumn{1}{l|}{S32 + L} 
    & \cellcolor{matteSkyblue25}87.24 
    & \cellcolor{matteSkyblue20}73.89 
    & \cellcolor{matteSkyblue25}82.16 
    & \cellcolor{matteSkyblue25}89.30 
    & \cellcolor{matteSkyblue25}75.11 
    & \cellcolor{matteSkyblue25}86.31 \\
\multicolumn{1}{l|}{S32 + M} 
    & \cellcolor{matteSkyblue25}85.30 
    & \cellcolor{matteSkyblue20}74.03 
    & \cellcolor{matteSkyblue25}82.31 
    & \cellcolor{matteSkyblue25}87.06 
    & \cellcolor{matteSkyblue25}75.51 
    & \cellcolor{matteSkyblue25}87.29 \\ 
\bottomrule
\end{tabular}
\caption{Performance Scores of different combinations; Cross-Attention is the baseline fusion technique and \textbf{\texttt{BAOMI}} is the proposed novel framework}
\label{tab:fusion}
\end{table}

\begin{figure}[!hbt]
    \centering
    \subfloat[]{%
        \includegraphics[width=0.23\textwidth]{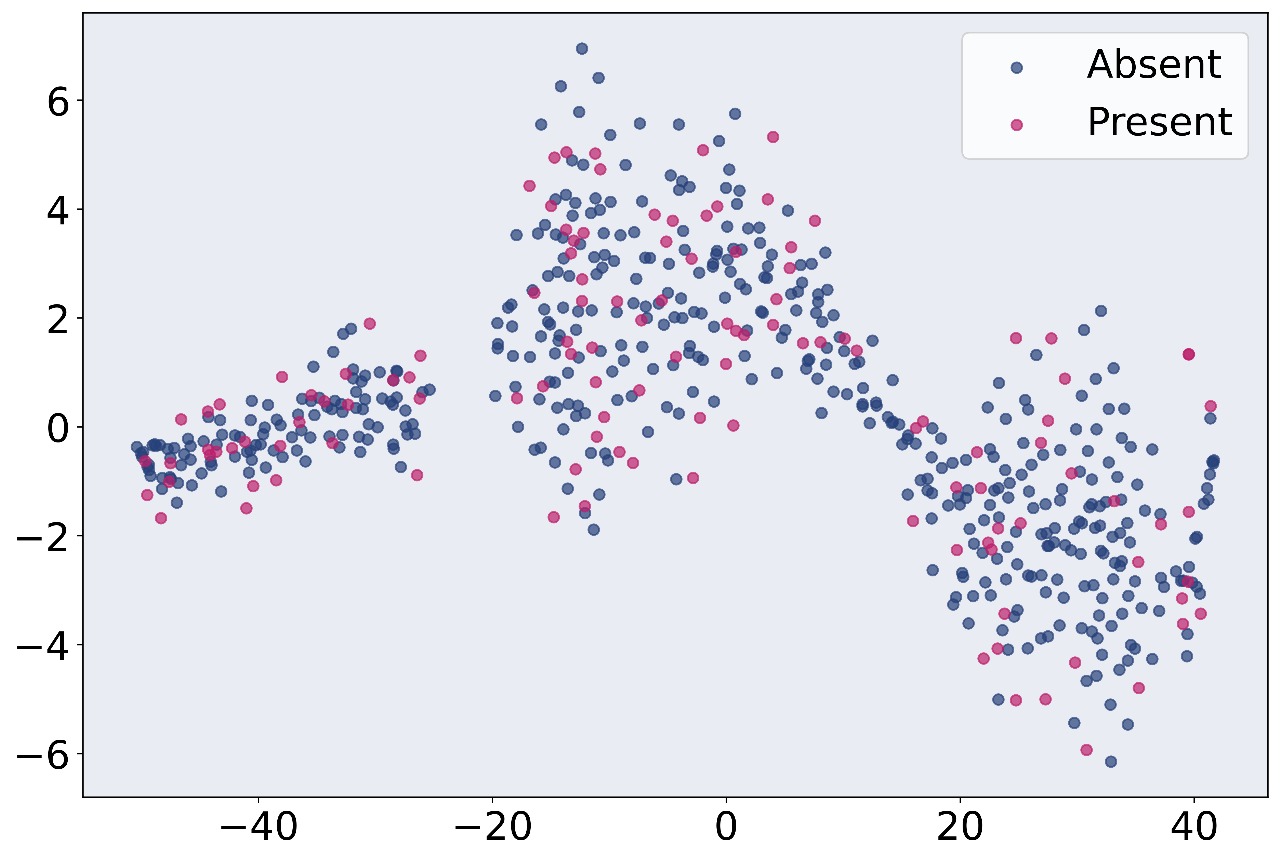}
    }
    \hfill
    \subfloat[]{%
        \includegraphics[width=0.23\textwidth]{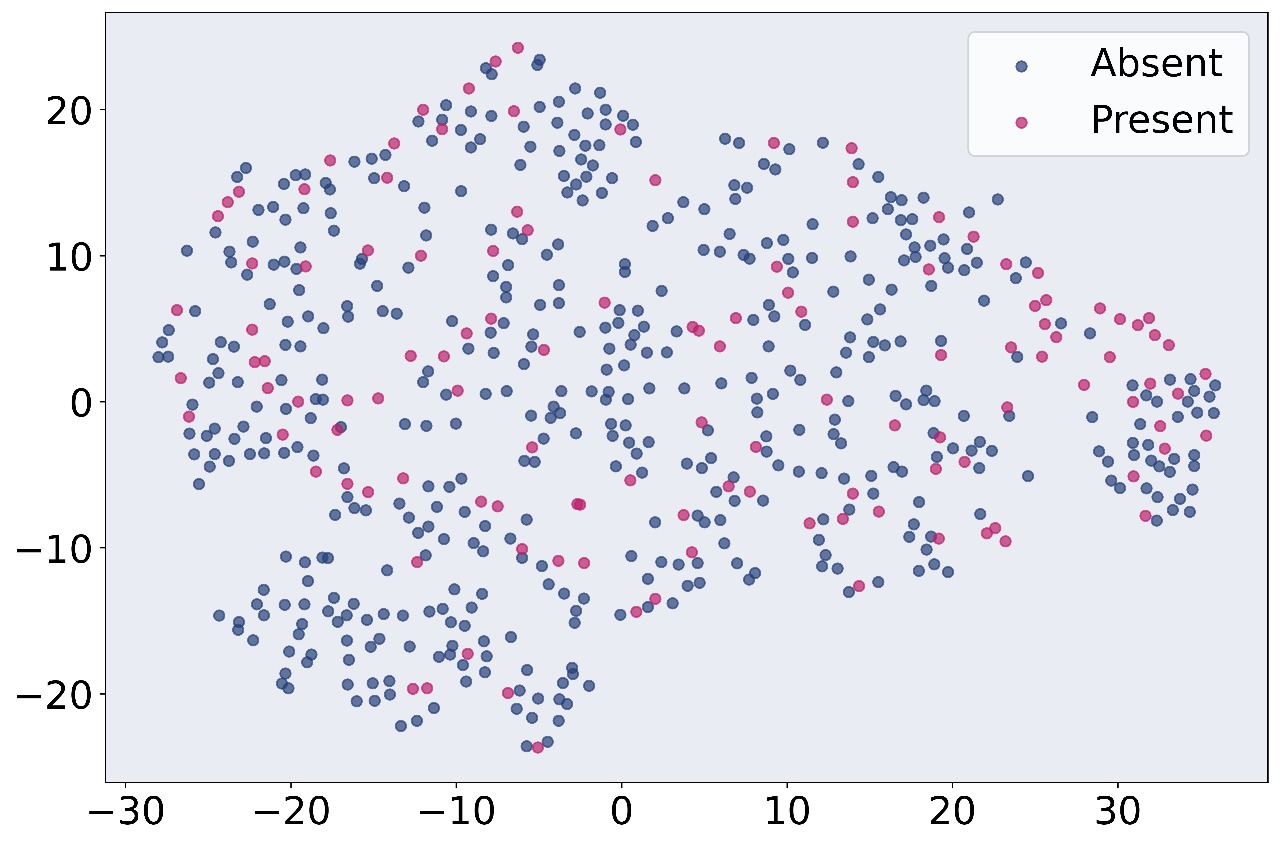}
    }
    \\
    \hfill
    \subfloat[]{%
        \includegraphics[width=0.23\textwidth]{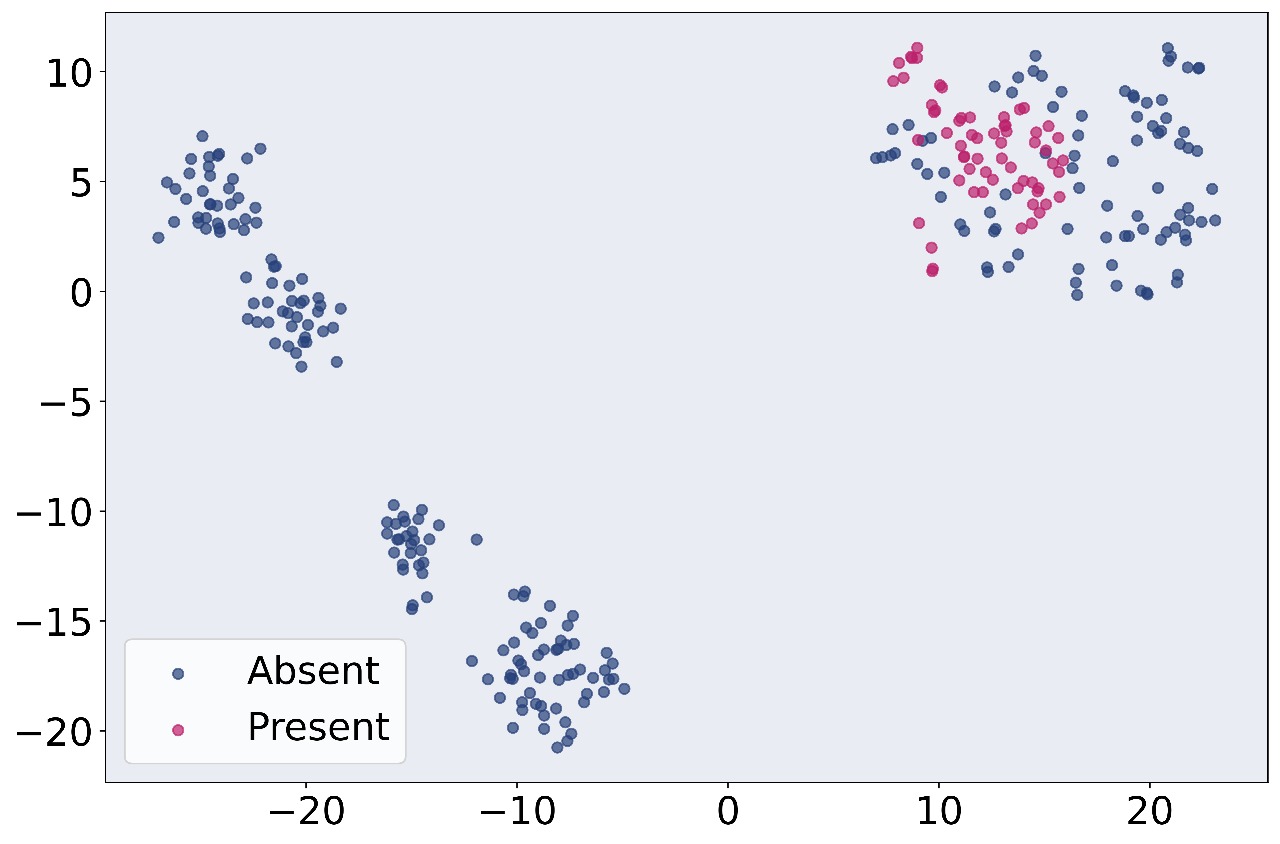}
    }
    \subfloat[]{%
        \includegraphics[width=0.23\textwidth]{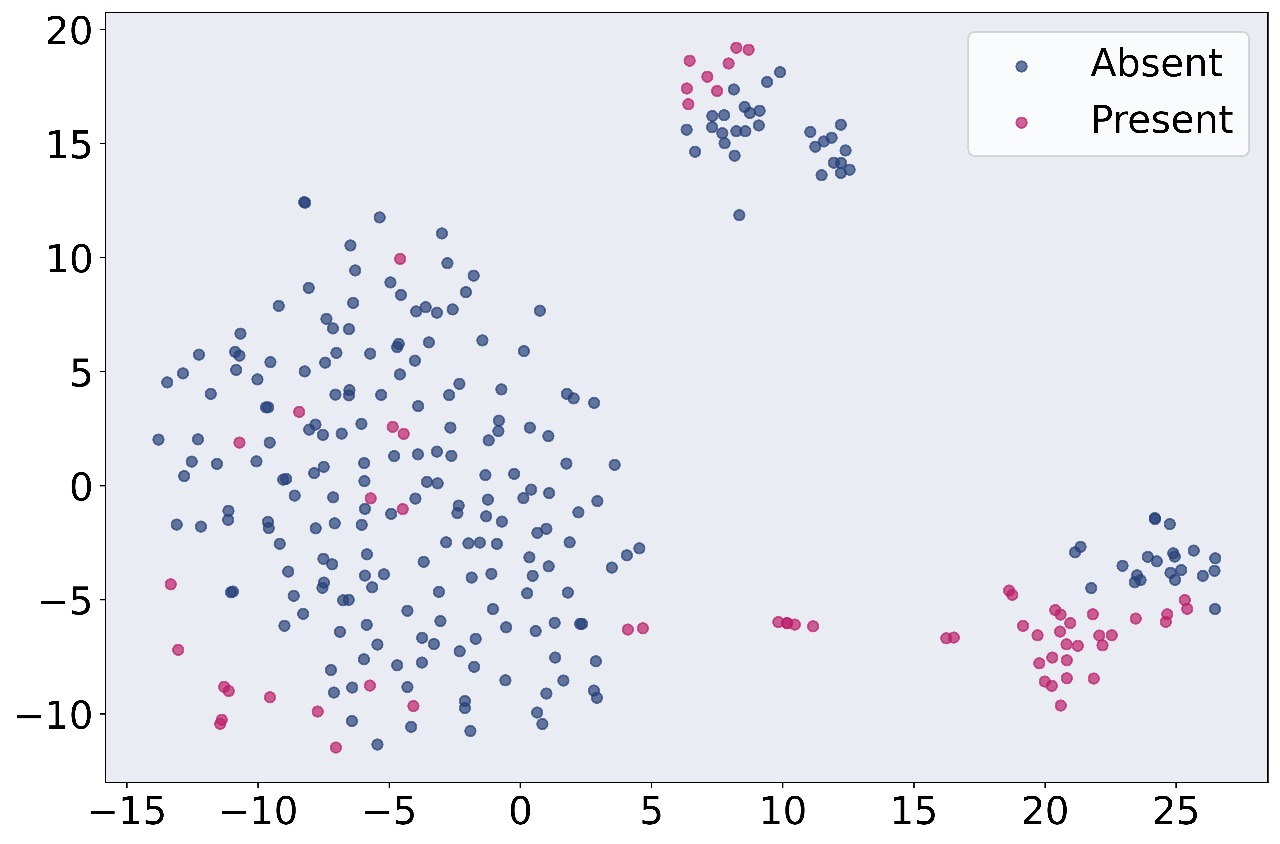}
    }
    \caption{t-SNE Plots- (a) CNN (MFCC) (b) CNN (LFCC) (c) \textbf{\texttt{BAOMI (DAC + MFCC)}}
    (d) \textbf{\texttt{BAOMI (SNAC24 + LFCC)}}}
    \label{fig:tsne}
\end{figure}

\begin{figure}[!h]
    \centering
    \subfloat[]{%
        \includegraphics[width=0.2\textwidth]{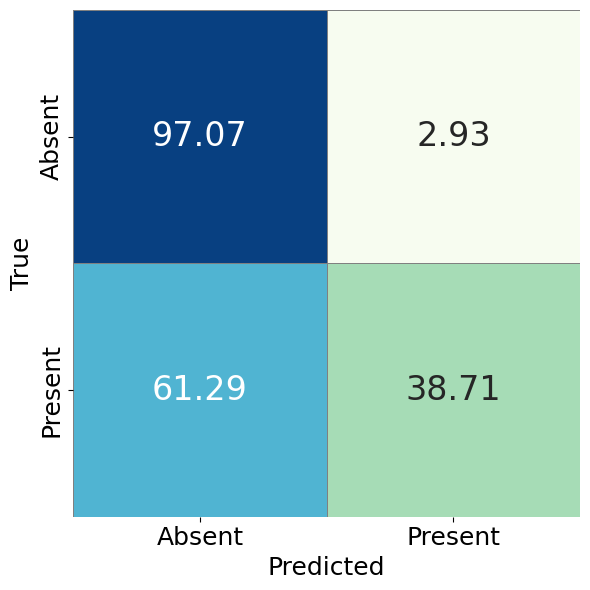}
    }
    \hfill
    \subfloat[]{%
        \includegraphics[width=0.2\textwidth]{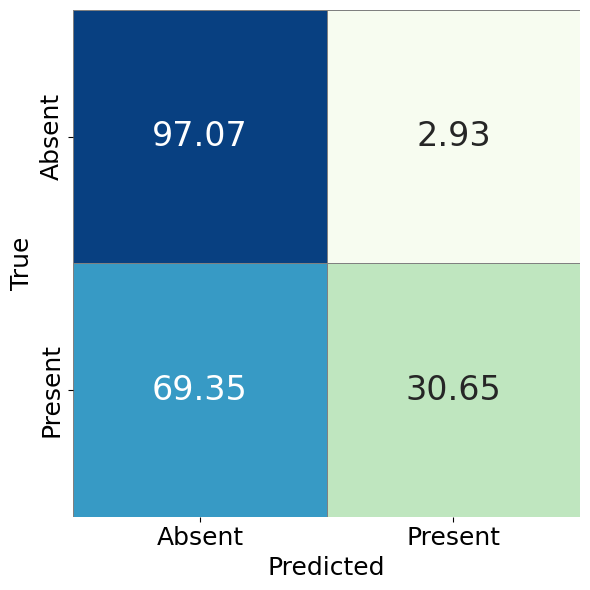}
    }
    \caption{Confusion Matrices - (a) \textbf{\texttt{BAOMI (DAC + MFCC)}} (b) \textbf{\texttt{BAOMI (SNAC24 + LFCC)}}}
    \label{fig:CM}
\end{figure}

\section{Conclusion}
In this work, we focus on HMC and hypothesize that combining NACRs with SFs will yield superior performance. To this end, we propose \textbf{\texttt{BAOMI}}, a novel framework that employs a bandit-based cross-attention mechanism to effectively fuse NACRs and SFs. By prioritizing the most important attention heads, \textbf{\texttt{BAOMI}} mitigates noise and enhances the fusion process. Our approach sets new SOTA performance, outperforming individual NACRs, SFs, and strong baseline fusion techniques. Our study will act as a guide as well as reference for future researchers exploring heterogeneous fusion of representations for improved HMC. The bandit-based cross-attention mechanism proposed in our work also find usage in different applications involving feature as well as multimodal fusion. 


\bibliographystyle{IEEEtran}
\bibliography{main}

\end{document}